\renewcommand{\Im}{\mathop{\rm Im}\nolimits}
\def\slash#1{\setbox0=\hbox{$#1$}               
   \dimen0=\wd0                                 
   \setbox1=\hbox{/} \dimen1=\wd1               
   \ifdim\dimen0>\dimen1                        
      \rlap{\hbox to \dimen0{\hfil/\hfil}}      
      #1                                        
   \else                                        
      \rlap{\hbox to \dimen1{\hfil$#1$\hfil}}   
      /                                         
   \fi}                                         %

\documentstyle[epsfig]{aipproc}

\begin{document}
\title{Gauge Invariance\\ and the\\ Unstable Particle}

\author{Robin G. Stuart}
\address{Randall Laboratory of Physics\\
Ann Arbor, Michigan 48109-1120\\
USA}

\maketitle

\begin{abstract}
It is shown how to construct exactly gauge-invariant $S$-matrix elements
for processes involving unstable gauge particles such as the $Z^0$
boson. The results are
applied to derive a physically meaningful expression
for the cross-section $\sigma(e^+e^-\rightarrow Z^0Z^0)$ and thereby
provide a solution to the long-standing
{\it problem of the unstable particle}.
\end{abstract}

\section{Introduction}

A resonance is fundamentally a non-perturbative object and is thus
not amenable to the methods of standard perturbation theory. In order to
describe physics at the $Z^0$ resonance one is forced, therefore, to employ
some sort of non-perturbative procedure. Such a procedure is Dyson
summation that sums strings of one-particle irreducible (1PI)
self-energy diagrams as a geometric series
to all orders in the coupling constant, $\alpha$
and effectively replaces the tree-level $Z^0$ propagator by a
dressed propagator,
\begin{equation}
\frac{1}{s-M_Z^2}
\rightarrow\frac{1}{s-M_Z^2}
    \sum_n\left(\frac{\Pi_{ZZ}(s)}{s-M_Z^2}\right)^n
=\frac{1}{s-M_Z^2-\Pi_{ZZ}^{(1)}(s)}
\label{eq:DysonSum}
\end{equation}
where $\Pi_{ZZ}(s)$ is the one-loop $Z^0$ self-energy.
The problem here is that electroweak physics is described by a gauge
theory. Results of calculations of physical processes must be exactly
gauge-invariant but this comes about through delicate cancellations
between many different Feynman diagrams each of which is separately
gauge-dependent. The
cancellation happens at each order in $\alpha$ when all diagrams of
a given order are combined. The $Z^0$ self-energy, $\Pi_{ZZ}(s)$,
is gauge-dependent at ${\cal O}(\alpha)$ and hence the rhs of
eq.(\ref{eq:DysonSum}) is gauge-dependent at all orders in $\alpha$.

If the dressed propagator is used in a finite-order calculation the
result will be gauge-dependent at some order because the will be no
diagrams available to cancel the gauge-dependence beyond the order
being calculated. This gauge-dependence should be
viewed as an indicator that the approximation scheme being used is
inconsistent or does not represent a physical observable.

In constructing amplitudes that represent physical observables, one must
take care to respect the requirements that are laid down by analytic
$S$-matrix theory \cite{AnalyticSMat}. These conditions are derived
from general considerations such as energy conservation and causality
and it will be found that appealing to them leads to a procedure for
generating gauge-invariant amplitudes with no flexibility in the final
result. The procedure will also make it possible to give an answer to
the {\it problem of the unstable particle\/} that was put forward
by Peierls in the early fifties \cite{Peierls}. Much of what appears
here can be found in ref.s\cite{Stuart1,Stuart3,Unstable}.

\section{Gauge-invariant $S$-matrix elements}

We will begin by reviewing what is known about $S$-matrix elements
for processes containing unstable particles. An unstable particle
is associated with a pole, $s_p$, lying on the second Riemann sheet
below the real $s$ axis. The scattering amplitude, $A(s)$ for a process
which contains an intermediate unstable particle can then be written in
the form,
\begin{equation}
A(s)=\frac{R}{s-s_p}+B(s)
\label{eq:AnalyticAmp}
\end{equation}
where $R$ and $s_p$ are complex constants and $B(s)$ is regular at
$s=s_p$. The first term on the rhs of eq.(\ref{eq:AnalyticAmp}) will
be called the resonant term and the second is the non-resonant background
term. It is known that $s_p$ is process-independent in the sense that
any process that contains a given unstable particle as an intermediate
state will have its pole at the same position. It can also be shown from
Fredholm theory that the residue factorizes as, $R=R_i\cdot R_f$, into
pieces that depend separately on the initial- and final-state.
One can prove, by very simple arguments, that $s_p$, $R$ and $B(s)$
are separately and exactly gauge-invariant \cite{Stuart1}.

>From analytic $S$-matrix theory it is known that production thresholds
are associated with a branch cut. Branch points for stable particles
lie on the real $s$-axis and those for unstable particles, such as $W$
bosons, lie significantly below it. Provided there are no nearby
thresholds the amplitude, $A(s)$, can be adequately described in the
resonance region by a Laurent expansion about the pole, $s_p$.
It should be emphasized that $A(s)$ can always be written in the form
(\ref{eq:AnalyticAmp}) even when thresholds are present. In that case
$B(s)$ will be an analytic function containing a branch point and the
rhs of (\ref{eq:AnalyticAmp}) continues to be an exact representation
of $A(s)$. Laurent expansion refers to how the resonant term is
identified. The aim is only to separate $A(s)$ into gauge-invariant
resonant and background pieces and there is no necessity to perform
a Laurent expansion beyond its leading term although this can provide
a useful way for parameterizing electroweak data \cite{SMatData}.

Let us consider the production process for a massless fermion pair in
$e^+e^-\rightarrow f\bar f$. Away from the $Z^0$ resonance it is known
how to calculate the amplitude to arbitrary accuracy
in a gauge-invariant manner. Near
resonance we must perform a Dyson summation and then separate the
amplitude into its resonant and background pieces. Doing so allows
expressions for the gauge-invariant quantities, $s_p$, $R$ and $B(s)$
to be identified in terms of the 1PI functions that occur in perturbation
theory. The scattering amplitude for $e^+e^-\rightarrow f\bar f$ is

\begin{eqnarray}
A(s,t)&=&{R_{iZ}(s_p)R_{Zf}(s_p)\over s-s_p}\nonumber\\
      &+&{R_{iZ}(s)R_{Zf}(s)-R_{iZ}(s_p)R_{Zf}(s_p)\over s-s_p}
       +{V_{i\gamma}(s)V_{\gamma f}(s)\over s-\Pi_{\gamma\gamma}(s)}
       +B(s,t)
\label{eq:ExactAmp}
\end{eqnarray}
in which
\begin{eqnarray}
R_{iZ}(s)&=&\left[V_{i\gamma}(s)
   \frac{\Pi_{\gamma Z}(s)}{s-\Pi_{\gamma\gamma}(s)}
                     +V_{iZ}(s)\right]F_{ZZ}^{\frac{1}{2}}(s),\\
R_{Zf}(s)&=&F_{ZZ}^{\frac{1}{2}}(s)\left[V_{Zf}(s)
       +\frac{\Pi_{Z\gamma}(s)}{s-\Pi_{\gamma\gamma}(s)}
                V_{\gamma f}(s)\right].
\label{eq:ResidueFactor}
\end{eqnarray}
The pole position $s_p$ is a solution of the equation
\begin{equation}
s-M_Z^2-\Pi_{ZZ}(s)
-{\Pi_{Z\gamma}^2(s)\over s-\Pi_{\gamma\gamma}(s)}=0
\label{eq:PoleEqn}
\end{equation}
and $F_{ZZ}(s)$ is defined through the relation
\begin{equation}
s-M_Z^2-\Pi_{ZZ}(s)
-{\Pi_{Z\gamma}^2(s)\over s-\Pi_{\gamma\gamma}(s)}
={1\over F_{ZZ}(s)}(s-s_p).
\label{eq:FZZdef}
\end{equation}

It should be emphasized that eq.(\ref{eq:ExactAmp}) is exact and valid anywhere
on the complex $s$-plane. The effect of $Z^0$-$\gamma$ mixing has been
included. The quantity $\Pi_{\gamma\gamma}(s)$ and $\Pi_{Z\gamma}(s)$ are
the photon self-energy and the $Z$-$\gamma$ mixing
respectively. $V_{iZ}(q^2)$, $V_{Zf}(q^2)$ are the
initial- and final-state $Z^0$ vertices, into which the external
wavefunctions have been absorbed, and
$V_{i\gamma}(q^2)$, $V_{\gamma f}(q^2)$ are the corresponding
vertices for the photon. Here $B(s,t)$ denotes 1PI corrections
to the matrix element that include things like as box diagrams.
The first term on the rhs of eq.(\ref{eq:ExactAmp}) is the resonant
part of $A(s,t)$ and the three terms on the second line taken together
are form the non-resonant background.

Calculations are most conveniently performed in terms of the real
renormalized parameters of the theory such as the renormalized mass,
$M_Z$. Eq.(\ref{eq:PoleEqn}) can be solved iteratively in terms of $M_Z$
to give
\begin{equation}
s_p=M_Z^2+\Pi_{ZZ}(M_Z^2)+...
\label{eq:PoleSoln}
\end{equation}
The rhs of eq.(\ref{eq:PoleSoln}) may be substituted for $s_p$ where it
appears in eq.(\ref{eq:ExactAmp})--(\ref{eq:ResidueFactor}). Taylor series
expansion can then be used to obtain perturbative expressions for $s_p$,
$R$ and $B(s)$ in terms of Greens functions with real arguments up to
any desired order. At any given order these perturbative expressions will
be exactly gauge-invariant as will scattering amplitudes constructed from
them.

A couple of points should be noted here.
The appearance of Greens functions with complex arguments
in eq.(\ref{eq:ExactAmp}) is a natural consequence
of the analyticity of the $S$-matrix. The $S$-matrix itself is never
evaluated with complex $s$. The arguments of $A(s,t)$ on the lhs of
eq.(\ref{eq:ExactAmp}) is real as is the `$s$' in the denominator of the
first term on the rhs.

In the procedure described above, one starts by extracting the resonant
term in a scattering amplitude by Laurent expansion about the exact pole
position $s_p$ and then specializes to lower orders by further expanding
about the renormalized mass. Other authors
\cite{AeppCuypOlde,AeppOldeWyl}
have attempted to apply the techniques described above by first expanded
about the renormalized mass and then added a finite width in the denominator
of the resonant part by hand. That procedure cannot be justified and leads
to problems when one treats processes like $e^+e^-\rightarrow W^+W^-$.
It gives rise to spurious {\it threshold singularities} or complex
scattering angles due the production threshold's being incorrectly
located on the real axis. In section IV the process
$e^+e^-\rightarrow Z^0Z^0$ will be treated and no threshold singularities
or complex scattering angles will arise.

\section{The Problem of the Unstable Particle}

We have thus succeeded in our goal of producing exactly
gauge-invariant scattering amplitudes to arbitrary order. One might ask
whether what has been done is just a mathematical trick, in which case
the gauge-invariance is accidental, or does it have some physical
interpretation. In this section it will be shown that the latter is true.

Recall that the coordinate space dressed propagator for a scalar
particle has an integral representation
\begin{equation}
\Delta(x^\prime-x)=\int \frac{d^4k}{(2\pi)^4}
                   \frac{e^{-ik\cdot(x^\prime-x)}}
                                 {k^2-m^2-\Pi(k^2)+i\epsilon}
\end{equation}
The integrand has a pole at $k^2=s_p$ where $s_p$ is a solution of the
equation $s-m^2+\Pi(s)=0$ and as in the previous section we define
$F(s)$ via the relation $s-m^2+\Pi(s)=(s-s_p)/F(s)$.
The dressed propagator can then be written as
\begin{equation}
\Delta(x^\prime-x)=\int \frac{d^4k}{(2\pi)^4}
        e^{-ik\cdot(x^\prime-x)}
         \left[\frac{F(s_p)}{k^2-s_p}
         +\frac{F(k^2)-F(s_p)}{k^2-s_p}\right]
\label{eq:splitprop}
\end{equation}
that separates resonant and non-resonant pieces. Performing the
$k_0$ integration resonant gives
\begin{eqnarray}
\Delta(x^\prime-x)&=&-i\int \frac{d^3k}{(2\pi)^3 2k_0}
         e^{-ik\cdot(x^\prime-x)}\theta(t^\prime-t)F(s_p)\nonumber\\
      & &+\int\frac{d^4k}{(2\pi)^4}\frac{F(k^2)-F(s_p)}{k^2-s_p}\\
      & &-i\int \frac{d^3k}{(2\pi)^3 2k_0}
         e^{ik\cdot(x^\prime-x)}\theta(t-t^\prime)F(s_p)\nonumber
\end{eqnarray}
where $k_0=\sqrt{\vec{k}^2+s_p}$.
The non-resonant term contributes only for $t=t^\prime$ and so
represents a contact interaction. The resonant part spits into two terms
that contribute when $t>t^\prime$ or $t<t^\prime$ and
therefore connects points $x$ and $x^\prime$ that are
separated by a finite distance in space-time.

The problem of the unstable particle \cite{Peierls} is may be roughly
stated as follows:
$S$-matrix theory deals with asymptotic in-states and out-states that
propagate from and to infinity. Unstable particles cannot exist as
asymptotic states because they decay a finite distance from the
interaction region. Indeed it is known \cite{Veltman} that the $S$-matrix
is unitary on the Hilbert space spanned by stable particle states and
hence there is not even any room to accommodate unstable particles as
external states. How can one use the $S$-matrix to calculate, say, the
production cross-section for an unstable particle when it cannot exist as
an asymptotic state?

In the first part of this section it was shown that the resonant part of
the dressed propagator connected points with a finite space-time
separation. When a similar analysis is applied to a physical matrix
element, such as eq.(\ref{eq:ExactAmp}), one concludes that the resonant
part describes a process in which there is a finite space-time separation
between the initial-state vertex, $V_i$, and the final-state vertex, $V_f$.
In other words, the resonant term describes the finite propagation of a
physical $Z^0$ boson. The non-resonant background represents prompt
production of the final state. As these two possibilities are, in principle,
physically distinguishable, they must be separately gauge-invariant.

We can thus use finite propagation as a tag for identifying unstable
particles without requiring that they appear in the final state.
This is, after all, the way $b$-quarks are identified in vertex detectors.
A production cross-section for an unstable particle is obtained by
extracting the resonant part of the matrix element for a process containing
that particle in an intermediate state and summing over all possible
decay modes.

\section{The process $\lowercase{e^+e^-}\rightarrow Z^0Z^0$}

In this section we will calculate the cross-section for
$e^+e^-\rightarrow Z^0Z^0$. This is of both theoretical and practical
importance. On the theoretical side it represents an example of a
calculation of the production cross-section for unstable particles.
On the practical side, at high energies $e^+e^-\rightarrow Z^0Z^0$
will be a dominant source of fermion pairs $(f_1\bar f_1)$ and
$(f_2\bar f_2)$ due to its double resonant enhancement and hence
$\sigma(e^+e^-\rightarrow Z^0Z^0)$ is an excellent approximation
to the cross-section for 4-fermion pair production.
If the experimental situation warrants it, background terms
can also be included without difficulty.

In the case of $e^+e^-\rightarrow f\bar f$, dealt with in section II,
the invariant mass squared of the $Z^0$, $s$,
is fixed by the momenta of the
incoming $e^+e^-$. For the process $e^+e^-\rightarrow Z^0Z^0$ the invariant
mass of the produced $Z^0$'s is not constant
and must be somehow included in phase
space integrations. It is not immediately clear how to do this and without
further guidance from $S$-matrix theory there would seem to be considerable
flexibility in how to proceed. A new ingredient is required and that is
to realize that an expression for an $S$-matrix element can always be
divided into a part that is a Lorentz-invariant function of the kinematic
invariants of the problem and Lorentz-covariant objects, such as $\slash{p}$
etc. The latter are known as {\it standard covariants\/}
\cite{Hearn,Hepp1,Williams,Hepp2}. It is the Lorentz invariant part that
satisfies the requirements of analytic
$S$-matrix theory and from which the resonant
and non-resonant background parts are extracted while the Lorentz
covariant part is untouched.

To calculate $\sigma(e^+e^-\rightarrow Z^0Z^0)$, we begin by constructing
the cross-section for
$e^+e^-\rightarrow Z^0Z^0\rightarrow (f_1\bar f_1)(f_2\bar f_2)$,
and will eventually sum over all fermion species.
The part of the full matrix element that can give rise to doubly
resonant contributions can be written as
\begin{eqnarray}
{\cal M}&=&\sum_i [\bar v_{e^+} T^i_{\mu\nu} u_{e^-}]
M_i(t,u,p_1^2,p_2^2)\nonumber\\
& &\ \ \ \ \ \ \ \times\frac{1}{p_1^2-M_Z^2-\Pi_{ZZ}(p_1^2)}
[\bar u_{f_1}\gamma^\mu(V_{Zf_L}(p_1^2)\gamma_L
                       +V_{Zf_R}(p_1^2)\gamma_R) v_{\bar f_1}]
\label{eq:fullZZ}\\
& &\ \ \ \ \ \ \ \times\frac{1}{p^2_2-M_Z^2-\Pi_{ZZ}(p_2^2)}
[\bar u_{f_2}\gamma^\nu(V_{Zf_L}(p_2^2)\gamma_L
                       +V_{Zf_R}(p_2^2)\gamma_R) v_{\bar f_2}]
\nonumber
\end{eqnarray}
where $T_{\mu\nu}^i$ are Lorentz covariant tensors that span the tensor
structure of the matrix element and $\gamma_L$, $\gamma_R$ are the
usual helicity projection operators.
The squared invariant masses of the $f_1\bar f_1$ and $f_2\bar f_2$ pairs
are $p_1^2$ and $p_2^2$. The $M_i$, $\Pi_{ZZ}$ and $V_{Zf}$
are Lorentz scalars that are analytic functions of the
independent kinematic Lorentz invariants of the problem.

To extract the piece of the matrix element that corresponds to finite
propagation of both $Z^0$'s we extract the leading term in a Laurent
expansion in $p_1^2$ and $p_2^2$ of the analytic Lorentz-invariant
part of eq.(\ref{eq:fullZZ}) leaving the Lorentz-covariant part
untouched. This is the doubly-resonant term and is given by
\begin{eqnarray}
{\cal M}&=&\sum_i [\bar v_{e^+} T^i_{\mu\nu} u_{e^-}]
M_i(t,u,s_p,s_p)\nonumber\\
& &\ \ \ \ \ \ \ \times\frac{F_{ZZ}(s_p)}{p_1^2-s_p}
[\bar u_{f_1}\gamma^\mu(V_{Zf_L}(s_p)\gamma_L
                       +V_{Zf_R}(s_p)\gamma_R) v_{\bar f_1}]
\label{eq:resZZ}\\
& &\ \ \ \ \ \ \ \times\frac{F_{ZZ}(s_p)}{p_2^2-s_p}
[\bar u_{f_2}\gamma^\nu(V_{Zf_L}(s_p)\gamma_L
                       +V_{Zf_R}(s_p)\gamma_R) v_{\bar f_2}]
\nonumber
\end{eqnarray}
where $F_{ZZ}$ defined by a relation like (\ref{eq:FZZdef}).
It should be emphasized that eq.(\ref{eq:resZZ}) is the exact form
of the doubly-resonant matrix element to all orders in perturbation
theory that we will now specialize to leading order.
It is free of threshold singularities noted that were found by other
authors \cite{AeppOldeWyl}.
In lowest order eq.(\ref{eq:resZZ}) becomes, up to overall multiplicative
factors,
\begin{eqnarray}
{\cal M}&=&\sum_{i=1}^2 [\bar v_{e^+} T^i_{\mu\nu} u_{e^-}]M_i\nonumber\\
& &\ \ \ \ \times\frac{1}{p_1^2-s_p}
[\bar u_{f_1}\gamma^\mu(V_{Zf_L}\gamma_L+V_{Zf_R}\gamma_R)v_{\bar f_1}]
\label{eq:lowestZZ}\\
& &\ \ \ \ \times\frac{1}{p_2^2-s_p}
[\bar u_{f_2}\gamma^\nu(V_{Zf_L}\gamma_L+V_{Zf_R}\gamma_R)v_{\bar f_2}].
\nonumber
\end{eqnarray}
where
$T^1_{\mu\nu}=\gamma_\mu(\slash{p}_{e^-}-\slash{p}_1)\gamma_\nu$,
$M_1=t^{-1}$;
$T^2_{\mu\nu}=\gamma_\nu(\slash{p}_{e^-}-\slash{p}_2)\gamma_\mu$,
$M_2=u^{-1}$
and the final state vertex corrections take the form
$V_{Zf_L}=ie\beta_L^f\gamma_L$ and $V_{Zf_R}=ie\beta_R^f\gamma_R$,
The left- and right-handed couplings of the $Z^0$ to a fermion $f$ are
\[
\beta_L^f=\frac{t_3^f-\sin^2\theta_W Q^f}{\sin\theta_W\cos\theta_W},
\hbox to 2cm{}
\beta_R^f=-\frac{\sin\theta_W Q^f}{\cos\theta_W}.
\]

Squaring the matrix element and integrating over the final state
momenta for fixed $p_1^2$ and $p_2^2$ gives
\begin{eqnarray}
\frac{\partial^3\sigma}{\partial t\,\partial p_1^2\,\partial p_2^2}
                   &=&\frac{\pi\alpha^2}{s^2}
                    (\vert\beta_L^e\vert^4+\vert\beta_R^e\vert^4)
                    \rho(p_1^2)\ \rho(p_2^2)\label{eq:diffxsec}\\
                   &\times&\left\{\frac{t}{u}+\frac{u}{t}
                     +\frac{2(p_1^2+p_2^2)^2}{ut}
                     -p_1^2p_2^2\left(\frac{1}{t^2}+\frac{1}{u^2}\right)
                     \right\}
\nonumber
\end{eqnarray}
with
\begin{eqnarray*}
\rho(p^2)&=&\frac{\alpha}{6\pi}
    \sum_f(\vert\beta_L^f\vert^2+\vert\beta_R^f\vert^2)
          \frac{p^2}{\vert p^2-s_p\vert^2}
               \theta(p_0)\theta(p^2)\\
 &\approx&\frac{1}{\pi}
         .\frac{p^2 (\Gamma_Z/M_Z)}{(p^2-M_Z^2)^2+\Gamma_Z^2 M_Z^2}
          \theta(p_0)\theta(p^2)
\end{eqnarray*}
where the sum is over fermion species.
Note that $\rho(p^2)\rightarrow \delta(p^2-M_Z^2)\theta(p_0)$
as $\Im(s_p)\rightarrow 0$ which is the result obtained by cutting a
free propagator. The variables $s$, $t$, $u$, $p_1^2$ and $p_2$ in
eq.(\ref{eq:diffxsec}) arise from products of standard covariants and external
wave functions and therefore take real values dictated by the kinematics.

Integrating over $t$, $p_1^2$ and $p_2^2$ leads to
\begin{equation}
\sigma(s)=\int_0^s dp_1^2
          \int_0^{(\sqrt{s}-\sqrt{p_1^2})^2} dp_2^2
          \sigma(s;p_1^2,p_2^2)\ \rho(p_1^2)\ \rho(p_2^2),
\label{eq:ZZxsec}
\end{equation}
where
\begin{eqnarray*}
\sigma(s;p_1^2,p_2^2)&=&\frac{2\pi\alpha^2}{s^2}
       (\vert\beta_L^e\vert^4+\vert\beta_R^e\vert^4)\\
&\times&
       \left\{\left(\frac{1+(p_1^2+p_2^2)^2/s^2}
{1-(p_1^2+p_2^2)/s}\right)\ln\left(\frac{-s+p_1^2+p_2^2+\lambda}
                                    {-s+p_1^2+p_2^2-\lambda}\right)
-\frac{\lambda}{s}\right\}
\end{eqnarray*}
and $\lambda=\sqrt{s^2+p_1^4+p_2^4-2sp_1^2-2sp_2^2-2p_1^2p_2^2}$.
For $p_1^2=p_2^2=M_Z^2$ this agrees with known results \cite{Brown}.

\end{document}